\documentclass[aps,prb,preprint,titlepage,groupedaddress,tightenlines,longbibliography,amsfonts,amssymb,amsmath]{revtex4-2}
\usepackage{lineno,hyperref,amsmath,graphicx}

\DeclareMathOperator{\sn}{sn}
\DeclareMathOperator{\sgn}{sgn}
\newtheorem{lemma}{Lemma}

\begin{document}

\title{Monte Carlo simulation of a model cuprate}
\author{Yu D Panov}
\author{A S Moskvin}
\author{A A Chikov}
\author{V A Ulitko}
\affiliation{Institute of Natural Sciences and Mathematics, Ural Federal University, 19 Mira str., Ekaterinburg , Russia }
\email[]{yuri.panov@urfu.ru}
\begin{abstract}
We develop a classical Monte Carlo algorithm based on a quasi-classical approximation for a pseudospin $S=1$ Hamiltonian in real space to construct a phase diagram of a model cuprate with a high T$_c$. 
A model description takes into account both local and nonlocal correlations, Heisenberg spin-exchange interaction, correlated single-particle, and two-particle transport. 
We formulate a state selection algorithm for a given parameterization of the wave function in order to ensure a uniform distribution of states in the phase space.
The simulation results show a qualitative agreement with the experimental phase diagrams. 
\end{abstract}
\maketitle
\section{Introduction}

One of the topical problems of the high-$T_c$ cuprate physics is the coexistence and competition of antiferromagnetic, superconducting, and charge orderings~\cite{Fradkin2012}. 
Recent accurate measurements of various physical characteristics on thousands of cuprate samples~\cite{Bozovic2016} indicate fundamental discrepancies with the ideas based on the canonical Bardeen-Cooper-Schrieffer approach, and rather support the bosonic mechanism of the high-$T_c$ cuprates.
The study is complicated by the presence of heterogeneity due to dopants or non-isovalent substitution, as well as to the internal electronic tendency to heterogeneity~\cite{Moskvin2019}.
A large number of theoretical models have been developed to account for the exotic electronic properties of cuprates in the normal state and to reveal the nature of unconventional superconductivity. 
However, to date, the problem of a consistent theoretical description of the cuprate phase diagram is still far from being solved. 

Earlier, we developed a minimal model of cuprates where the CuO$_2$ planes are considered as lattices of CuO$_4$ clusters, which are the main element of the crystal and electronic structure of cuprates. 
The on-site Hilbert space~\cite{Moskvin2011,Moskvin2013} is formed by three effective valence states of the cluster: [CuO$_4$]$^{7-}$, [CuO$_4$]$^{6-}$, and [CuO$_4$]$^{5-}$. 
The very possibility of considering these centers on an equal basis is due to the strong electron-lattice relaxation effects in cuprates~\cite{Mallett2013,Moskvin2019-2}. 
The centers differ in the usual spin state: $s=1/2$ for the [CuO$_4$]$^{6-}$ center and $s=0$ for the [CuO$_4$]$^{7-}$ and [CuO$_4$]$^{5-}$, respectively, and in the orbital symmetry: $B_{1g}$ for the ground states of the [CuO$_4$]$^{6-}$ center, $A_{1g}$ for the [CuO$_4$]$^{7-}$ center, and the Zhang-Rice $A_{1g}$ or more complicated low-lying non-Zhang-Rice states for the [CuO$_4$]$^{5-}$ center. 
In these many-electron atomic complexes with strong $p{-}d$ covalence and strong intra-center correlations, electrons cannot be described within any conventional (quasi)particle approach that addresses the [CuO$_4$]$^{7-,6-,5-}$ centers within the on-site hole representation $|n\rangle$, 
$n = 0, 1, 2$, respectively. 
Instead of conventional quasiparticle $k$-momentum description, 
we make use of a real space on-site $S=1$ pseudospin formalism to describe the charge triplets and an effective spin-pseudospin Hamiltonian which takes into account both local and nonlocal correlations, single and two-particle transport, as well as Heisenberg spin-exchange interaction. 

The pseudospin approach is used for the strongly correlated electron systems~\cite{Castellani1979,Rice1981} 
and for the superconductivity~\cite{Low1994} of cuprates for a long time. 
The pseudospin formalism leads to the possibility of simulation using the well-developed classical Monte Carlo (MC) method for constructing phase diagrams and studying the features of the thermodynamic properties of the system. 
A similar effective $S=1$ spin-charge model for cuprates and its MC implementation 
were considered in recent papers~\cite{Cannas2019,Frantz2021}.

We organize the article as follows. 
In Section 2, we present the $S=1$ pseudospin formalism and the effective spin-pseudospin Hamiltonian of the model. 
In Section 3, we introduce quasi-classical approximation for the wave functions and formulate the state selection algorithm.
The results of classical MC simulations of our model and their discussion are presented in Section 4.

\section{Model}
A minimal model to describe the charge degree of freedom in cuprates~\cite{Moskvin2011,Moskvin2013}
implies that for the CuO$_4$ centers in CuO$_2$ plane
the on-site Hilbert space reduced to a charge triplet
formed by the three many-electron valence states [CuO]$_4^{7-,6-,5-}$ (nominally Cu$^{1+, 2+, 3+}$).
These states can be considered to be the components of the $S=1$ pseudospin triplet with projections $M_S = {-}1,\, 0,\, {+}1$.
Effective pseudospin Hamiltonian of the model cuprate with the addition of the Heisenberg spin-spin exchange coupling
of the $s=1/2$ [CuO]$_4^{6-}$ (Cu$^{2+}$) centers can be written as follows:
\begin{equation}
		\mathcal{H}
		= \mathcal{H}_{ch} + \mathcal{H}_{ex}
		+ \mathcal{H}_{tr}^{(1)}
		+ \mathcal{H}_{tr}^{(2)}
		- \mu \sum_i S_{zi} .
		\label{eq:Ham0}
\end{equation}
Here, the first term
\begin{equation}
	\mathcal{H}_{ch} =
	\Delta \sum_i S_{zi}^2
	+ V \sum_{\left\langle ij\right\rangle} S_{zi} S_{zj}
\end{equation}
describes the on-site and inter-site nearest-neighbour density-density correlations, respectively,
so that $\Delta=U/2$, $U$ being the correlation parameter, and $V>0$.
The sums run over the sites of a 2D square lattice, $\left\langle ij \right\rangle$ means the nearest neighbors.
The second term
\begin{equation}
		\mathcal{H}_{ex}
		=
		Js^2 \sum_{\langle ij \rangle} \boldsymbol{\sigma}_i \boldsymbol{\sigma}_j
\end{equation}
is the  antiferromagnetic ($J>0$) Heisenberg exchange coupling for the CuO$_4^{6-}$ centers,
where $\boldsymbol{\sigma}=P_0 \mathbf{s}/s$ operators take into account the on-site spin density $P_0 = 1-S_z^2$,
and $\mathbf{s}$ is the spin $s=1/2$ operator.
The third term has the following form:
\begin{multline}
	\mathcal{H}_{tr}^{(1)}
	\;=\;
	- t_p \sum_{\left\langle ij\right\rangle }
		\big(  P_{i}^{+} P_{j}^{}  +   P_{j}^{+} P_{i}^{} \big)
	- t_n \sum_{\left\langle ij\right\rangle }
		\big(  N_{i}^{+} N_{j}^{}  +  N_{j}^{+} N_{i}^{}  \big)
	 \\
	{} - \frac{t_{pn}}{2}
	\sum_{\left\langle ij\right\rangle }
		\big(  P_{i}^{+} N_{j}^{}  +  P_{j}^{+} N_{i}^{}
		+ N_{i}^{+} P_{j}^{}  +  N_{j}^{+} P_{i}^{}  \big) , 
\end{multline}
where the transfer integrals $t_p$, $t_n$, $t_{pn}$
describe three types of the correlated "one-particle" transport.
$P$ and $N$ operators are the combinations of the pseudospin $S=1$ operators~\cite{Moskvin2011}:
$P^{+} = \tfrac{1}{2} \left(S_{+} + T_{+}\right)$, 
$N^{+} = \tfrac{1}{2} \left(S_{+} - T_{+}\right)$, 
where $T_{+} = S_z S_{+} + S_{+} S_z$.
The next term is
\begin{equation}
	\mathcal{H}_{tr}^{(2)}
	= - t_b \sum_{\left\langle ij\right\rangle} \big(  S_{{+}i}^2 S_{{-}j}^2 + S_{{+}j}^2 S_{{-}i}^2  
	\big) ,
\end{equation}
where the transfer integral $t_b$ describes the two-particle ("composite boson") transport
\cite{Moskvin2011}.
The last term with the chemical potential $\mu$ allows to account the constraint for the total charge density $n$: 
\begin{equation}
	n = \frac{1}{N}\left\langle \sum_i S_{zi} \right\rangle = const.
	\label{eq:charge_const}
\end{equation}

The mean-field approximation (MFA) for a model with the Hamiltonian \eqref{eq:Ham0} allowed us to find~\cite{Panov2019} the critical temperatures equations for the antiferromagnetic (AFM) ordering, charge ordering (CO), superconducting ordering (SC), and for the ``metal'' phase (M). 
The MFA phase diagram~\cite{Moskvin2020,Moskvin2021} of the model \eqref{eq:Ham0} demonstrates the possibility of correctly describing the features of phase diagrams typical of cuprates.

In the quasi-classical approximation, we write the on-site wave function of the charge triplet as follows
\begin{equation}
	\left| \Psi \right\rangle
	= c_{{+}1} \left| {+}1 \right\rangle 
	+ c_0 \left| 0 \right\rangle \left| \sigma \right\rangle 
	+ c_{{-}1} \left| {-}1 \right\rangle 	,
	\label{eq:Psi}
\end{equation}
where $\left| {+}1 \right\rangle$ and $\left| {-}1 \right\rangle$ are the orbital wave functions of [CuO$_4$]$^{5-}$ and [CuO$_4$]$^{7-}$ centers, respectively, 
$\left| 0 \right\rangle$ is the orbital wave function of [CuO$_4$]$^{6-}$ center, 
$\left| \sigma \right\rangle$ is the spin $s=1/2$ function:
\begin{equation}
	\left| \sigma \right\rangle
	=	e^{ -i \frac{\chi}{2} } \cos \frac{\eta}{2} \left| \uparrow \right\rangle
	+ e^{ i \frac{\chi}{2} } \sin \frac{\eta}{2} \left| \downarrow \right\rangle , 
	\label{eq:sigma}
\end{equation}
and the coefficients can be written in the following form:
\begin{equation}
	c_{{+}1} = e^{- i\frac{\alpha}{2}} \sin\frac{\theta}{2}\cos\frac{\phi}{2}  ,\quad
	c_{0} = e^{i\frac{\beta}{2}} \cos\frac{\theta}{2} , \quad
	c_{{-}1} = e^{ i\frac{\alpha}{2}} \sin\frac{\theta}{2}\sin\frac{\phi}{2} .
	\label{eq:ccc}
\end{equation}
Here $0 \le \theta \le \pi$,  $0 \le \phi \le \pi$, $0 \le \alpha \le 2\pi$, $0 \le \beta \le 2\pi$.

The matrices of pseudospin operators $S_z$ and $S_{\pm}$ in a basis 
$\left\{ \left| {+}1 \right\rangle , \left| 0 \right\rangle , \left| {-}1 \right\rangle \right\}$ are written as 
\begin{equation}
	S_z = 
	\begin{pmatrix}
	1 & 0 & 0 \\
	0 & 0 & 0 \\
	0 & 0 & -1 
	\end{pmatrix}
	,\quad
	S_{+} = 
	\begin{pmatrix}
	0 & 1 & 0 \\
	0 & 0 & 1 \\
	0 & 0 & 0 
	\end{pmatrix}
	,\quad
	S_{-} =
	\begin{pmatrix}
	0 & 0 & 0 \\
	1 & 0 & 0 \\
	0 & 1 & 0 
	\end{pmatrix}
	.
\end{equation}
Using equations (\ref{eq:Psi}-\ref{eq:ccc}) we can write average values for all operators in the Hamiltonian~\eqref{eq:Ham0} on the $i$th site:
\begin{eqnarray}
	\left\langle \Psi_i \left| S_{zi} \right| \Psi_i \right\rangle 
	&=& \frac{1}{2} \left( 1- \cos\theta_i \right) \cos\phi_i , \\
	\left\langle \Psi_i \left| S_{zi}^2 \right| \Psi_i \right\rangle 
	&=& \frac{1}{2} \left( 1- \cos\theta_i \right) , \\
	\left\langle \Psi_i \left| S_{+ i}^2 \right| \Psi_i \right\rangle 
	&=& \frac{1}{4} \, e^{ i \alpha_i} \left( 1- \cos\theta_i \right) \sin\phi_i , \\
	\left\langle \Psi_i \left| P_{i}^{+} \right| \Psi_i \right\rangle 
	&=& \frac{1}{2} \, e^{i \frac{\alpha_i+\beta_i}{2}} \sin\theta_i \cos\frac{\phi_i}{2} , \\
	\left\langle \Psi_i \left| N_{i}^{+} \right| \Psi_i \right\rangle 
	&=& \frac{1}{2} \, e^{i \frac{\alpha_i-\beta_i}{2}} \sin\theta_i \sin\frac{\phi_i}{2} , \\
	\left\langle \Psi_i \left| \boldsymbol{\sigma}_i \right| \Psi_i \right\rangle
	&=& \frac{1}{2} \left( 1 + \cos\theta_i \right) 
	\big\{ \sin \eta_i  \cos \chi_i , \, \sin \eta_i \sin \chi_i , \, \cos \eta_i  \big\} , 
\end{eqnarray}
where the $x$-, $y$-, and $z$-components of vector are listed in curly brackets. 
This allows us to obtain the energy for a model~\eqref{eq:Ham0} in the quasi-classical approximation
\begin{equation}
	E = \Big\langle \prod_i \Psi_i \Big| \, \mathcal{H} \, \Big| \prod_i \Psi_i \Big\rangle
\end{equation}
in the following form:
\begin{eqnarray}
	E 
	&=& 
	\frac{\Delta}{2} \sum_i \left(1 - \cos\theta_i\right) 
	+ \frac{V}{4} \sum_{\left\langle ij\right\rangle}  
	\left(1 - \cos\theta_i\right)  \left(1 - \cos\theta_j\right) \cos\phi_i \, \cos\phi_j  
	\nonumber\\[0.5em]
	&&{} 
	+\frac{J}{16} \sum_{\langle ij \rangle} 
	\left( 1 + \cos\theta_i \right)  \left( 1 + \cos\theta_j \right)
	\big(  \sin \eta_i \sin \eta_j \cos \left( \chi_i-\chi_j \right) + \cos \eta_i \cos \eta_j  \big)
	\nonumber\\[0.5em]
	&&{}
	- \frac{t_p}{2} \sum_{\left\langle ij\right\rangle} 
	\sin\theta_i \, \sin\theta_j \, \cos\frac{\phi_i}{2} \, \cos\frac{\phi_j}{2} \, 
	\cos \frac{\alpha_i-\alpha_j+\beta_i-\beta_j}{2}
	\nonumber\\[0.5em]
	&&{}
	-\; \frac{t_n}{2} \sum_{\left\langle ij\right\rangle} 
	\sin\theta_i \, \sin\theta_j \, \sin\frac{\phi_i}{2} \, \sin\frac{\phi_j}{2} \, 
	\cos \frac{\alpha_i-\alpha_j-\beta_i+\beta_j}{2}
	\nonumber\\[0.5em]
	&&{} 
	- \frac{t_{pn}}{4} \,  \sum_{\left\langle ij\right\rangle} 
	\sin\theta_i \, \sin\theta_j \, 
	\bigg( 
	\sin\frac{\phi_i+\phi_j}{2} \,  \cos \frac{\alpha_i-\alpha_j}{2} \,  \cos \frac{\beta_i+\beta_j}{2}
	\nonumber\\[0.5em]
	&&{}\qquad
	+ \sin\frac{\phi_i-\phi_j}{2} \, \sin \frac{\alpha_i-\alpha_j}{2} \,  \sin \frac{\beta_i+\beta_j}{2}
	\bigg)
	\nonumber\\[0.5em]
	&&{}
	- \frac{t_B}{8} \sum_{\left\langle ij\right\rangle} 
	\left(1 - \cos\theta_i\right)  \left(1 - \cos\theta_j\right) \sin\phi_i \, \sin\phi_j \, \cos(\alpha_i-\alpha_j)
	\nonumber\\[0.5em]
	&&{}
	- \frac{\mu}{2} \sum_i \left(1 - \cos\theta_i\right) \cos\phi_i
	.
\end{eqnarray}
The constraint \eqref{eq:charge_const} can be written as:
\begin{equation}
	n = \frac{1}{2N} \sum_i \left(1 - \cos\theta_i\right) \cos\phi_i = const.
\end{equation}

\section{State selection algorithm}

\begin{figure}[t]
\centerline{\includegraphics[width=0.9\textwidth]{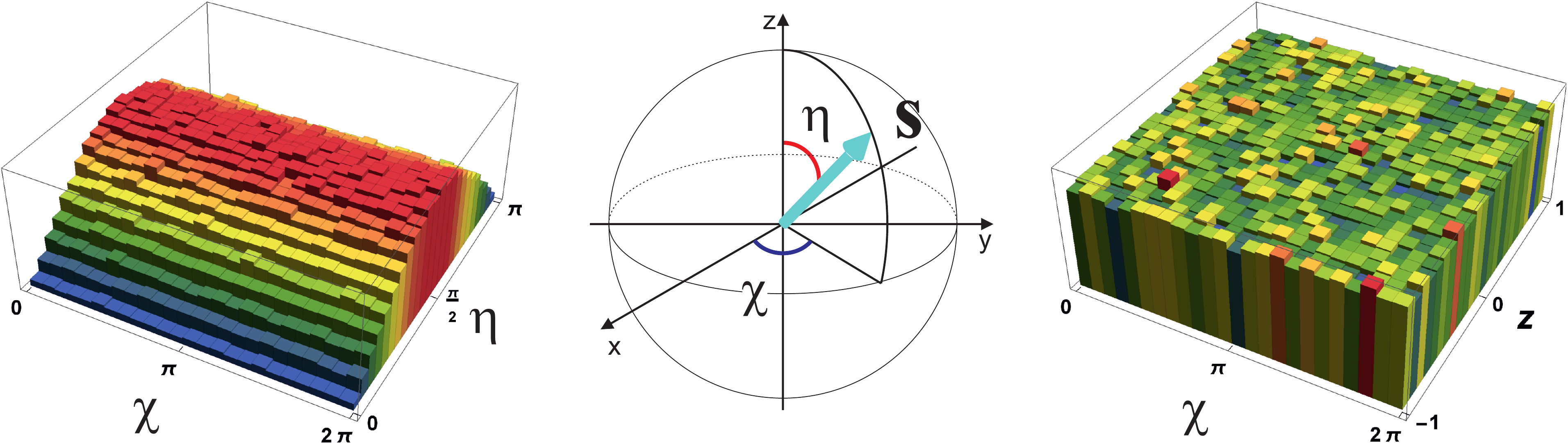}}
\caption{(Color online) The $(\chi,\eta)$-histogram (left panel), $(\chi,z)$-histogram (right panel)}
\label{fig:spin-hist}
\end{figure}

An average value of the spin $s=1/2$ operator
\begin{equation}
	\left\langle \sigma | \mathbf{s} | \sigma \right\rangle
	= \frac{1}{2}  \left\{ \sin \eta  \cos \chi , \, \sin \eta \sin \chi , \, \cos \eta \right\} 
\end{equation}
maps the spin states to the unit sphere. 
It is well known, that the uniform distribution of randomly generated points 
over the unit sphere is given by the following state selection algorithm:
\begin{enumerate}
	\item $\chi = 2 \pi \gamma_1$;
	\item $ \eta = \arccos  \gamma_2$, 
\end{enumerate}
where $\gamma_{1,2}$ are random numbers in the $[0,1]$ range. 
The $(\chi,\eta)$-histogram is shown in Fig.\ref{fig:spin-hist}, left panel, and this produces the flat $(\chi,z)$-histogram shown in Fig.\ref{fig:spin-hist}, right panel.

This state selection algorithm is based on the well-known lemmas~\cite{Sobol1973} of probability theory:

\begin{lemma}
Let $\gamma \in [0,1]$ is the uniformly distributed random value,
and $f(x)$ is the probability density function. 
Then the random value $\xi$ that satisfies the equation
$$\int_{-\infty}^{\xi} f(x)dx = \gamma$$
has the probability density function $f(x)$.
\end{lemma}

\begin{lemma}
Let $\gamma_1$ and $\gamma_2$ be uniformly distributed in $[0,1]$ random values, 
and $f(x_1,x_2)=f_1(x_1)f_2(x_2|x_1)$ is the joint density function,
where $f_1(x_1)$ is the marginal density, 
and $f_2(x_2|x_1)$ is the conditional density. 
Let $\xi_1$ and $\xi_2$ be continuous random variables that satisfy the system of equations
$$
\int_{-\infty}^{\xi_1} f_1(x)dx = \gamma_1, \qquad 
\int_{-\infty}^{\xi_2} f_2(x|\xi_1)dx = \gamma_2.
$$
Then $\xi_1$ and $\xi_2$ have the joint density function $f(x_1,x_2)$.
\end{lemma}

The states with coefficients~\eqref{eq:ccc} corresponds to a point in the octant of the unit sphere.
We use the Metropolis algorithm for a system with conservation of the total charge.
The charge at the site, $n_i$, is related to the parameters of the wave function by the expression
\begin{equation}
	2 n_i = \left( 1 - \cos \theta_i \right) \cos \phi_i .
\end{equation}
We require when the states of sites 1 and 2 change simultaneously, the total charge of the pair is preserved,
$n_1 + n_2 = n_1' + n_2' = 2n$, and the points representing states uniformly fill the allowed area in the octant.

The state selection algorithm based on the Lemmas 1 and 2 consists of the following steps:
\begin{enumerate}
	\item caclulation of $n_1$, where $-1+n+|n| \le n_1 \le 1+n-|n|$, from equation
	\begin{equation}
		G_1(n_1;n) = \gamma		
		,
	\end{equation}
	where $\gamma \in [0,1]$ is the uniformly distributed random value,
	\begin{equation}
		G_1(n_1;n) = \frac{ \Phi(n_1) - \Theta(n) \, \Phi(-1+2|n|) }{ \Phi(1-2|n|) }
		,
	\end{equation}
	\begin{equation}
		\Phi(x) =	\sgn x
		\Bigg[
		\frac{2\sqrt{1+|x|}}{\pi}
		\bigg(
		\frac{ 2 \, \Pi\left(-1,\frac{\pi}{2} \,\big|\, m(x) \right) }{ 1+|x| }
		-  m(x) \, K \left( m(x) \right)
		\bigg)
		-\frac{1}{2}
		\Bigg]
		+ \frac{1}{2} ,
	\end{equation}
	$m(x) = \left(1 - |x|\right) /\left(1 + |x|\right)$,
	$\Theta(x)$ is the Heaviside step function,
	$\Pi\left(-1,\frac{\pi}{2} \,\big|\, m \right) = \Pi_1 (1 , \sqrt{m}  )$ is the complete elliptic integral of the third kind, $K(m)$ is the complete elliptic integral of the first kind;

	\item calculation of the value $n_2 = 2n - n_1$;
	
	\item  calculation of $ \cos \frac{\theta_i}{2} $ from equation
	\begin{equation}
	\cos \frac{\theta_i}{2} = \sqrt{1 - |n_i|} \; \sn \left( \gamma_i K \left(m(n_i)\right) , m(n_i) \right)
	,
	\end{equation}
	where $\gamma_i \in [0,1]$, $i=1,2$, are the uniformly distributed random values, 
	$\sn \left( x , m \right)$ is the Jacobi function. 
	If $n_i=0$, we take $\cos \frac{\theta_i}{2} = \gamma_i$.
	
	\item calculation of $\cos \phi_i$ from equation
	\begin{equation}
		\cos \phi_i = \frac{n_i}{1-\cos^2 \frac{\theta_i}{2}}
		.
	\end{equation}
	If $n_i=0$ and $\cos \frac{\theta_i}{2}=1$, $\phi_i$ is a random uniformly distributed quantity, $0 \le \phi_i \le \pi$.
\end{enumerate}

\begin{figure}[t]
\centerline{\includegraphics[width=0.35\textwidth]{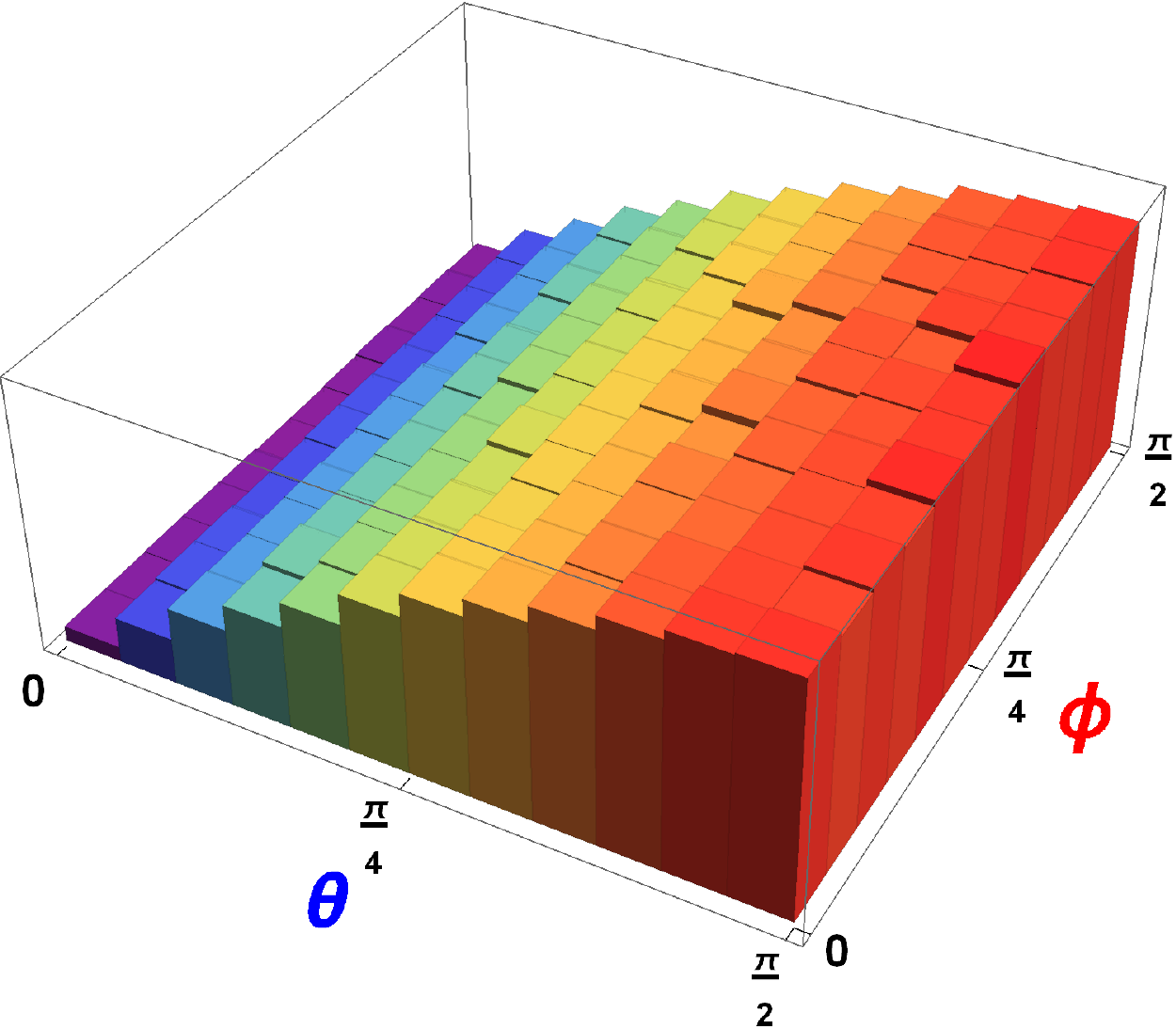} \qquad
\includegraphics[width=0.35\textwidth]{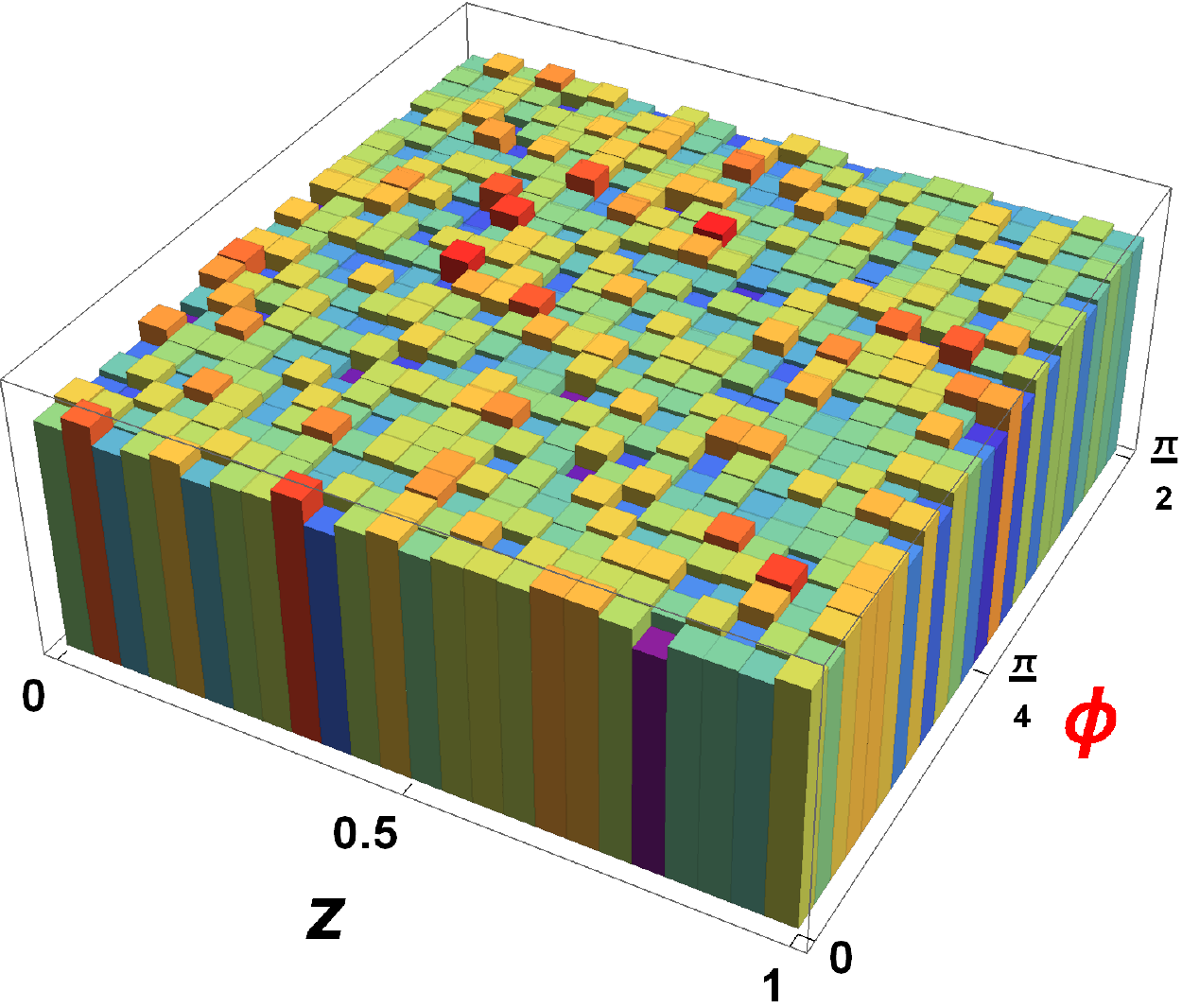} }
\caption{(Color online) The $(\phi,\theta)$-histogram (left panel), $(\phi,z)$-histogram (right panel)}
\label{fig:orb-hist}
\end{figure}

The distributions in the case $n=0$ for angles $(\phi,\theta)$ and for states on the octant of unit sphere are shown in Fig.\ref{fig:orb-hist}, left and right panels, respectively.

\section{Results}

\begin{figure}[t]
\centerline{\includegraphics[width=\textwidth]{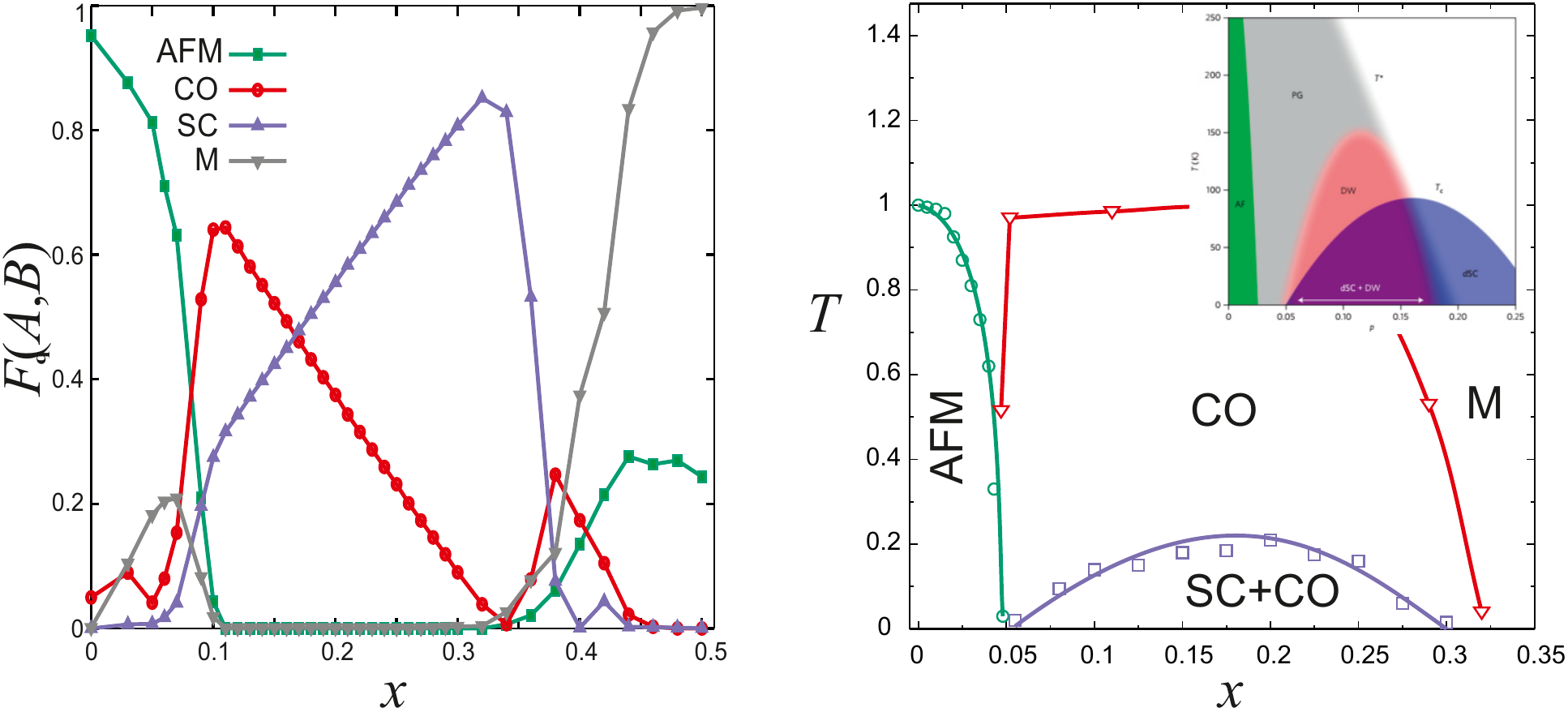}}
\caption{(Color online) Left panel: dependencies on the charge doping of the structure factors near the ground state calculated with parameters
$\Delta$\,=\,0.8, $V$\,=\,0.625, $J$\,=\,1, $t_p$\,=\,0.35, $t_n$\,=\,0, $t_{pn}$\,=\,-0.24, (all in units of the $t_b$). 
Right panel: The $T$\,-\,$x$ ($x$ the charge doping) phase diagram for the model cuprate calculated with the same parameters as in left panel. 
The insert shows a schematic phase diagram of hole-doped cuprates~\cite{Hamidian2016}.}
\label{fig3}
\end{figure}

In MC simulation, we calculated the structure factors
\begin{equation}
	F_{\mathbf{q}}(A,B) = \frac{1}{N^2} \sum_{lm} e^{i\mathbf{q}\,(\mathbf{r}_l - \mathbf{r}_m)}
	\left\langle A_{l} B_{m} \right\rangle
	,
\end{equation}
where $A_{l}$ and $B_{m}$ are the on-site operators and the summation is performed over all sites of the square lattice.
To determine the type of ordering, we monitored the following structure factors:
\begin{itemize}
	\item 
	$F_{(\pi,\pi)}(\boldsymbol{\sigma},\boldsymbol{\sigma})$ for antiferromagnetic (AFM) order,
	\item 
	$F_{(\pi,\pi)}(S_z,S_z)$ for the charge order (CO),
	\item 
	$F_{(0,0)}(S_{{+}}^2,S_{{-}}^2)$ for the superconducting order (SC),
	\item 
	$F_{(0,0)}(P^{+},P)$ for the ``metal'' phase ($M$).
\end{itemize}

The results of the MC simulation for the doping dependencies of the main structure factors near the ground state, $T/t_b=0.05$, of the model cuprate are presented in Fig.~\ref{fig3}, left panel. 
The critical temperatures for the AFM, CO, and SC phases were determined from the jump in the structure factor from zero to a certain finite value.
Fig.~\ref{fig3}, right panel, shows the reconstruction from the MC simulations of the $T$\,-\,$x$ phase diagram. 
The insert shows the  typical for the hole doped cuprates~\cite{Hamidian2016}. 

The obtained phase diagram for model cuprate with the Hamiltonian \eqref{eq:Ham0} reproduces some most important features of real phase diagrams: with given parameters, $\Delta=0.8$, $V=0.625$, $J=1$, $t_p=0.35$, $t_n=0$, $t_{pn}=-0.24$, (all in units of the $t_b$), near the "parent" composition $x=0$, we get AFM ordering, which is replaced with increasing $x$ by the SC ordering, that coexists with the CO ordering in the form of phase separation. The found SC ordering exits in the finite region of doping, and it is replaced by the ``metal'' phase at $x\geq0.3$. 
The main problems in the implementation of our modeling, such as inhomogeneous phase states and the associated difficulties in identifying them, are predetermined by the enhanced role of fluctuations in low-dimensional systems. 
But at the same time, the obtained phase diagrams show promising possibilities to describe the coexistence and competition of various phase orders in cuprates.

\section*{Acknowledgments}
The research was supported  by the Ministry of Education and Science of the Russian Federation, project FEUZ-2020-0054, and by scholarship of the president of the Russian Federation No. SP-2278.2019.1.

\providecommand{\newblock}{}

\end{document}